 \documentclass [twocolumn] {revtex4-2}

\usepackage{CJKutf8}
\usepackage {amsmath}
\usepackage {bbm}
\usepackage {color}
\usepackage {xcolor}
\usepackage{textpos}
\usepackage {physics}
\usepackage[compat=1.1.0]{tikz-feynman}
\usepackage {float}
\usepackage[makeroom]{cancel}
\makeatletter
\newcommand{\subalign}[1]{%
  \vcenter{%
    \Let@ \restore@math@cr \default@tag
    \baselineskip\fontdimen10 \scriptfont\tw@
    \advance\baselineskip\fontdimen12 \scriptfont\tw@
    \lineskip\thr@@\fontdimen8 \scriptfont\thr@@
    \lineskiplimit\lineskip
    \ialign{\hfil$\m@th\scriptstyle##$&$\m@th\scriptstyle{}##$\hfil\crcr
      #1\crcr
    }%
  }%
}
\usepackage {tikz}
\usetikzlibrary {datavisualization}
\usetikzlibrary {decorations.pathmorphing}

\makeatother

\begin {document}

	\title {The optical Bloch equation for the finite-temperature fluctuations}
	\author {\begin{CJK*}{UTF8}{gbsn}Weitao Liu	(刘\ 伟涛)\end{CJK*}}
	\affiliation {School of Physics, Nankai University, 300071 Tianjin, China}
	\begin {abstract}

		In this work, I analyze the quantum fluctuations and the thermal fluctuations in the framework of quantum mechanics. Being recognized as incoherent perturbations with different features, fluctuations of these two types lead to dissipative terms in the optical Bloch equations. The method allows one to use the optical Bloch equation to analyze time-dependent processes in the finite-temperature fluctuations. The numerical results show that the deexcitation is the limit of the equilibration at zero temperature. The impact of the fluctuations on the coherent excitations are also discussed.

	 \end {abstract}

	\maketitle

	\section {Introduction}
	
		Leave your hot cup of coffee or cold beer alone for a while and they soon reach the room temperature \cite{PhysRevE.79.061103}. Such phenomenon is called the equilibration. Equilibrations happen commonly in both the classical and the quantum regime. However, unlike in the classical regime, it is difficult to obtain a comprehensive understanding of the equilibrations in the quantum regime. This is because the principles of quantum mechanics usually lead to the time-reversal symmetry, while the equilibrations has the irreversibility. Some efforts have been made to explain the equilibration. In the Weisskopf-Wigner theory \cite{Weisskopf1930}, a Markovian approximation is introduced to describe the deexcitations, which is the equilibration at the zero temperature limit. In the Feynman-Vernon influence functional thoery \cite{FEYNMAN2000547}, the process is analyzed with the relation between the fluctuation and the dissipation \cite {PhysRev.32.110, PhysRev.90.977, phys.rev.83.34, PhysRev.101.1620}.
		
		The aim of this article is to provide a more straightforward method to include the equilibrations in the framework of quantum mechanics. Since the equilibration is the result of the quantum and the thermal fluctuation, which exist eternally in the free space, it impacts quantum systems commonly. During past decades, coherent perturbations, such as lasers, provide researchers opportunities to precisely control the quantum systems. Due to the fact that such artificial coherent perturbations are much more stronger than the fluctuations, in theoretical works, the impact of the fluctuations are usually omitted. However, technological improvements, such as the realization of the extremely long coherence time \cite{PhysRevLett.125.093201}, may make the impact of fluctuations prominent. Thus it is necessary to discuss the coupling between the coherent excitation and the fluctuations. Currently, coherent excitations are investigated by solving the time-dependent Schr\"odinger equation (TDSE) \cite{PhysRevLett.133.152503,PhysRevC.111.054316}. To include the deexcitation, the optical Bloch equations are also applicable \cite{Wense2020a}. To include the effects of the finite temperature, a straightforward method is to change the dissipative term in the optical Bloch equations. 
		
		The fluctuations have been investigated for more than one century. The earlist research dates back to Einstein and Ehrenfest \cite {Einstein1923}, who classified the emissions into the spontaneous ones and the stimulated ones, while all the absorptions are stimulated. The quantum fluctuation-dissipation theorem \cite {PhysRev.32.110, PhysRev.90.977, phys.rev.83.34, PhysRev.101.1620} reveals that in the equilibrium, the energy of oscillation with frequency $\omega$ is
		\begin {equation}
			E (\omega, \beta) = \frac{\omega}{2} +  \frac{\omega}{e^{\beta\omega}-1}.
			\label {eq:energy_mode}
		\end {equation}
		Herein and after we use the natural units with $\hbar=c=k=1$. $\beta=1/T$ is the inverse of the temperature. Eq. \eqref{eq:energy_mode} reveals the correspondence between the transitions and the fluctuations. The first term ${\omega}/{2}$ is energy of the quantum fluctuation, corresponding to the spontaneous emissions. Where the denominator $2$ comes from the fact that the spontaneous absorptions do not exist. The second term is the energy of the thermal fluctuation, corresponding to the stimulated transitions, which contain both stimulated emissions and stimulated absorptions. The thermal fluctuations obey the Bose-Einstein distribution. Since the equilibrium is the kept by the quantum fluctuation and the thermal fluctuation together, the question arises that whether these two types of fluctuations lead the equilibration when the system is out of the equilibrium. In this article, at first, I assume the answer to be true. Then the mathematical expressions of the fluctuations of these 2 types are constructed in the framework of quantum mechanics. At last, numerical realization of the mathematical expressions are used to check the assumption. 
		
		This article is organized as follows. The theoretical framework is derived in sec. \ref{sec:theory}; The numerical results from the theoretical framework are demonstrated in sec. \ref{sec:results}; Features of this method and more possible applications are discussed in sec. \ref{sec:conclusion}; A brief review of the Weisskopf-Wigner theory is provided in Appendix \ref{sec:W-W-theory}; A useful technique in the integration is attached in Appendix \ref{sec:integration}; The detailed derivation of the quantum fluctuations is shown in Appendix \ref{sec:quantum_fluctuation}.
		
	\section {Theoretical framework}
		\label {sec:theory} 
		
		The equilibration is an irreversible process of a series quantum transitions caused by the fluctuations. However, although the quantum transition is well studied with both perturbative and non-perturbative methods, direct use of these theoretical tools fails to explain the equilibration process. This is because the perturbative methods destroy the unitarity, leading to an unphysical total probability other than $1$; while the non-perturbative methods lead to periodic results, which violate the irreversibility. In the present work, perturbative and non-perturbative methods are combined to describe the impact of the fluctuations. For simplicity, a two-level system is discussed as a toy model. 
			
		\subsection {A quantum 2-level system subject to incoherent perturbations}
			\label {subsec:perturbation}
		
			When a 2-level system is subject to a perturbation with the frequency $\omega$, its evolution is described by the TDSE
			\begin {equation}
				i\frac{\partial}{\partial{t}} |\psi(t,\omega)\rangle = \left[H_0 + V(t,\omega)\right] |\psi(t,\omega)\rangle,
			\end {equation}
			where the unperturbed Hamiltonian
			\begin {equation}
				H_0 = 
				\begin {pmatrix}
					E_1 & 0 \\
					0 & E_2
				\end {pmatrix}
			\end {equation}
			determines the eigenstates of the system while the external perturbation
			\begin {equation}
				V(t,\omega) = 
				\begin {pmatrix}
					0 & ce^{-i\omega{t}} \\
					c^*e^{i\omega{t}} & 0
				\end {pmatrix}
				\label {eq:incoherent_perturbation}
			\end {equation}
			is responsible for the transitions. With the ansatz
			\begin {equation}
				|\psi (t,\omega)\rangle = a_1(t,\omega)e^{-iE_1t}|\psi_1\rangle + a_2(t,\omega)e^{-iE_2t}|\psi_2\rangle,
			\end {equation}
			it is straightforward to gets the first-order solution
			\begin {subequations}
				\begin {align}
					& \begin {aligned}
						a_1(t+\Delta{t},\omega) &= a_1(t) + a_2(t)c  \\ &\hspace{-10ex}\times\frac{e^{-i(E_2+\omega-E_1)(t+\Delta{t})}-e^{-i(E_2+\omega-E_1)t}}{-i(E_2+\omega-E_1)} \\
					\end {aligned}, \\
					& \begin {aligned}
						a_2(t+\Delta{t},\omega) &= a_2(t) - a_1(t)c^* \\ &\hspace{-10ex}\times \frac{e^{-i(E_1-\omega-E_2)(t+\Delta{t})}-e^{-i(E_1-\omega-E_2)t}}{-i(E_1- \omega-E_2)}. \\
					\end {aligned}
				\end {align}
				\label {eq:time_shift}
			\end {subequations}
			With the definition
			\begin {equation}
				\begin {aligned}
					A (\Delta{t},t,\omega,\epsilon) &= \\ &\hspace{-10ex} - c \frac{e^{-i(E_2+\omega-E_1)(t+\Delta{t})}-e^{-i(E_2+\omega-E_1)t}}{E_2+\omega-E_1+i\epsilon},
				\end {aligned}
				\label {eq:def_A}
			\end {equation}
			Eq. \eqref{eq:time_shift} is rewritten as
			\begin {equation}
				\begin {aligned}
					& \begin {pmatrix}
						a_1(t+\Delta{t},\omega,\epsilon) \\
						a_2(t+\Delta{t},\omega,\epsilon)
					\end {pmatrix}
					= \begin {pmatrix}
						a_1(t) \\
						a_2(t)
					\end {pmatrix}	
					\\ & \hspace{2ex} - i
					\begin {pmatrix}
						0 & A(\Delta{t}, t, \omega, \epsilon) \\
						A^*(\Delta{t}, t, \omega, \epsilon) & 0
					\end {pmatrix} 
					\begin {pmatrix}
						a_1(t) \\
						a_2(t)
					\end {pmatrix}.	
				\end {aligned}
				\label {eq:wave_transformation}
			\end {equation}
			In Eq. \eqref{eq:def_A}, an imaginary infinitesimal is added to the denominator. This is to avoid the divergence where $\omega=E_1-E_2$. Eq. \eqref {eq:time_shift} - eq. \eqref {eq:wave_transformation} provide a good approximation when  the transitions of the first order play the dominant role in $\Delta{t}$. To allow back and forth transitions between the two levels, eq. \eqref{eq:wave_transformation} is replaced by
			\begin {equation}
				\begin {aligned}
					&\begin {pmatrix}
						a_1(t+\Delta{t},\omega,\epsilon) \\
						a_2(t+\Delta{t},\omega,\epsilon)
					\end {pmatrix} \\
					= &\exp\left[ - i
					\begin {pmatrix}
						0 & A(\Delta{t},t,\omega,\epsilon) \\
						A^*(\Delta{t},t,\omega,\epsilon) & 0
					\end {pmatrix} \right] \\
					&\hspace{2em}\times
					\begin {pmatrix}
						a_1(t) \\
						a_2(t)
					\end {pmatrix}.
				\end {aligned}
				\label {eq:single_transition}
			\end {equation}
			With the definition of the frequency-independent density matrix
			\begin {equation}
				\rho(t) = 
				\begin {pmatrix}
					a_1(t)a_1^*(t) & a_1(t)a_2^*(t) \\
					a_2(t)a_1^*(t) & a_2(t)a_2^*(t)
				\end {pmatrix}
				\label {def:density_matrix}
			\end {equation}
			and the frequency-dependent transfer matrix
			\begin {equation}
				\begin {aligned}
				&T(\Delta{t},t,\omega,\epsilon)\\
				&\hspace{5ex} = \exp\left[ - i
				\begin {pmatrix}
					0 & A(\Delta{t},t,\omega,\epsilon) \\
					A^*(\Delta{t},t,\omega,\epsilon) & 0
				\end {pmatrix} \right],
				\end {aligned}
				\label {eq:transfer_matrix}
			\end {equation}
			the frequency dependent density matrix, defined as
			\begin {equation}
				\begin {aligned}
				&\rho(t,\omega,\epsilon) \\
				&\hspace{3ex}= 
				\begin {pmatrix}
					a_1(t,\omega,\epsilon)a_1^*(t,\omega,\epsilon) & a_1(t,\omega.\epsilon)a_2^*(t,\omega,\epsilon) \\
					a_2(t,\omega,\epsilon)a_1^*(t,\omega,\epsilon) & a_2(t,\omega,\epsilon)a_2^*(t,\omega,\epsilon) \\
				\end {pmatrix},
				\end {aligned}
			\end {equation}
			is obtained by
			\begin {equation}
				\rho(t+\Delta{t},\omega,\epsilon) = T(\Delta{t},t,\omega,\epsilon) \rho(t) T^\dagger(\Delta{t},t,\omega,\epsilon).
				\label {eq:density_evolution}
			\end {equation}
			It should be noted that with the phase factor dropped, the density matrices defined throughout this article is slightly different from textbooks on quantum mechanics. This difference does not affect the physical results. One may note that Eq. \eqref {def:density_matrix} set the restriction that the system is in a pure state at the time $t$. However, for a two-level system, one can always decompose the mixed state into two pure states. Thus the following derivations are valid when $\rho(t)$ represent a mixed state. It can be derived that, up to the second order of $A(\Delta{t},t,\omega,\epsilon)$, the elements of $\rho(t+\Delta{t},\omega,\epsilon)$ are
			\begin {subequations}
				\begin {align}
					& \begin {aligned}
						\rho_{11}(t+\Delta{t},\omega,\epsilon) =&\left(1-|A(\Delta{t},t,\omega,\epsilon)|^2\right)\rho_{11}(t) \\ 
						& + iA^*(\Delta{t},t,\omega,\epsilon)\rho_{12}(t) \\
						& - iA(\Delta{t},t,\omega,\epsilon)\rho_{21}(t)  \\
						& + |A(\Delta{t},t,\omega,\epsilon)|^2\rho_{22}(t),
					\end {aligned} \\
					& \begin {aligned}
						 \rho_{12}(t+\Delta{t},\omega,\epsilon) = &\ {iA(\Delta{t},t,\omega,\epsilon)}\rho_{11}(t) \\
						 & + \left(1-|A(\Delta{t},t,\omega,\epsilon)|^2\right)\rho_{12}(t) \\
						 & + {A^2(\Delta{t},t,\omega,\epsilon)}\rho_{21}(t) \\
						 & - {iA(\Delta{t},t,\omega,\epsilon)}\rho_{22}(t),
					\end {aligned} \\
					& 
					\begin {aligned}
						\rho_{21}(t+\Delta{t},\omega,\epsilon) = & - {iA^*(\Delta{t},t,\omega,\epsilon)}\rho_{11}(t) \\
						& + {{A^*}^2(\Delta{t},t,\omega,\epsilon)}\rho_{12}(t) \\
						& + (1-|A(\Delta{t},t,\omega,\epsilon)|^2)\rho_{21}(t) \\
						& + {iA^*(\Delta{t},t,\omega,\epsilon)}\rho_{22}(t),
					\end {aligned} \\
					& 
					\begin {aligned}
						\rho_{22}(t+\Delta{t},\omega,\epsilon) = & |A(\Delta{t},t,\omega,\epsilon)|^2\rho_{11}(t) \\
						& - iA^*(\Delta{t},t,\omega,\epsilon)\rho_{12}(t) \\
						& + iA(\Delta{t},t,\omega,\epsilon)\rho_{21}(t) \\
						& + (1 - |A(\Delta{t},t,\omega,\epsilon)|^2)\rho_{22}(t).
					\end {aligned}
				\end {align}
				\label {eq:monochromatic_transformation}
			\end {subequations}
			Eq. \eqref{eq:monochromatic_transformation} shows that during the time $[t,t+\Delta{t})$, the monochromatic perturbation $V(t,\omega)$ makes a shift $\Delta\rho(\Delta{t},t,\omega,\epsilon)$ on the density matrix $\rho(t)$, with the elements
			\begin {subequations}
				\begin {align}
					& \begin {aligned}
						\Delta\rho_{11}(\Delta{t},t,\omega,\epsilon) = & -|A(\Delta{t},t,\omega,\epsilon)|^2\rho_{11}(t) \\
						& + iA^*(\Delta{t},t,\omega,\epsilon)\rho_{12}(t) \\
						& - iA(\Delta{t},t,\omega,\epsilon)\rho_{21}(t) \\
						& + |A(\Delta{t},t,\omega,\epsilon)|^2\rho_{22}(t),
					\end {aligned} \\
					& \begin {aligned}
						\Delta\rho_{12}(\Delta{t},t,\omega,\epsilon) = &\ {iA(\Delta{t},t,\omega,\epsilon)}\rho_{11}(t) \\
						& - |A(\Delta{t},t,\omega,\epsilon)|^2\rho_{12}(t) \\
						& + {A^2(\Delta{t},t,\omega,\epsilon)}\rho_{21}(t) \\
						& - {iA(\Delta{t},t,\omega,\epsilon)}\rho_{22}(t),
					\end {aligned} \\
					& 
					\begin {aligned}
						\Delta\rho_{21}(\Delta{t},t,\omega,\epsilon) = & - {iA^*(\Delta{t},t,\omega,\epsilon)}\rho_{11}(t) \\
						& + {{A^*}^2(\Delta{t},t,\omega,\epsilon)}\rho_{12}(t) \\
						& - |A(\Delta{t},t,\omega,\epsilon)|^2\rho_{21}(t) \\
						& + {iA^*(\Delta{t},t,\omega,\epsilon)}\rho_{22}(t),
					\end {aligned} \\
					& 
					\begin {aligned}
						\Delta\rho_{22}(\Delta{t},t,\omega,\epsilon) = &\ |A(\Delta{t},t,\omega,\epsilon)|^2\rho_{11}(t) \\
						& - iA^*(\Delta{t},t,\omega,\epsilon)\rho_{12}(t) \\
						& + iA(\Delta{t},t,\omega,\epsilon)\rho_{21}(t) \\
						& - |A(\Delta{t},t,\omega,\epsilon)|^2\rho_{22}(t).
					\end {aligned}
				\end {align}
				\label {eq:monochromatic_shift}
			\end {subequations}
			An integration over the frequency space is necessary to evaluate the effects of incoherent perturbations. Assume in the frequency interval $d\omega$, the perturbation has the state density $n(\omega)$, then the number of perturbation modes is $d\omega n(\omega)$. The shift of the density matrix during $\Delta{t}$ is 
			\begin {equation}
				\begin {aligned}
					\Delta\rho(\Delta{t},t) = \lim_{\epsilon\rightarrow0^+} \int_{-\infty}^{+\infty}{d\omega}n(\omega)\Delta{\rho}(\Delta{t},t,\omega,\epsilon).
				\end {aligned}
			\end {equation}
			Using the residue theorem, one can prove that in the limit $\epsilon\rightarrow0^+$, in eq. \eqref{eq:monochromatic_shift}, only terms with the coefficient $|A(\Delta{t},t,\omega,\epsilon)|^2$ contribute to the result (see Appendix \ref{sec:integration}) . Consequently, 
			\begin {subequations}
				\begin {align}
					& \begin {aligned}
					\Delta{\rho}_{11}(\Delta{t},t) &= - 2\pi|c|^2n(\Delta{E})  \\
					&\qquad\times (\rho_{11}(t) - \rho_{22}(t)) \Delta{t}, \\
					\end {aligned} \\
					& \Delta{\rho}_{12}(\Delta{t},t) = - 2\pi|c|^2n(\Delta{E})\rho_{12}(t)\Delta{t}, \\
					& \Delta{\rho}_{21}(\Delta{t},t) = - 2\pi|c|^2n(\Delta{E})\rho_{21}(t)\Delta{t}, \\
					& \begin {aligned} 
					\Delta{\rho}_{22}(\Delta{t},t) &= - 2\pi|c|^2n(\Delta{E}) \\
					&\qquad\times(\rho_{22}(t) - \rho_{11}(t))\Delta{t},
					\end {aligned}
				\end {align}
				\label {eq:incoherent_shift}
			\end {subequations}
			where $\Delta{E}=E_2-E_1$. Eq. \eqref {eq:incoherent_shift} reveals that, although the term $|A(\Delta{t},t,\omega,\epsilon)|^2$ is of the order $\mathcal{O}(\Delta{t}^2)$, its integration over the frequency space is of the order $\mathcal{O}(\Delta{t})$. This difference reflects the uncertainty principle that in the time interval $\Delta{t}$,  perturbations of the frequencies within the width $2\pi/\Delta{t}$ contribute to the quantum system.
		
		\subsection {The quantum fluctuations}
			\label {subsec:quantum_fluctuation}
		
\begin {figure}
				\includegraphics {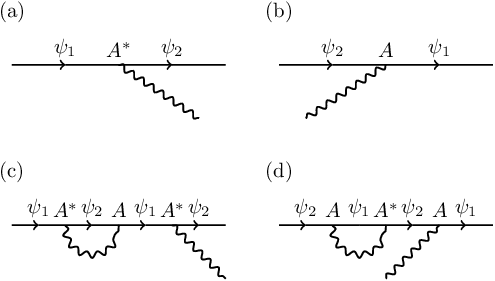}
				\caption {Feynman diagrams of one photon emissions and absorptions. Spontaneous emissions (a) and (c) are counted while spontaneous absorptions (b) and (d) are omitted.}
				\label {fig:feynman}
			\end {figure}

			In Eq. \eqref{eq:incoherent_shift}, the time derivatives of the diagonal elements are proportional to the difference between the diagonal elements. This leads to the final state that both the excited state and the ground state get the occupation probability $1/2$. However, this result does not agree with the realistic observations. In the case $\Delta{E}>0$, $|\psi_1\rangle$ is the excited state and $|\psi_2\rangle$ is the ground state. The system evolves to the equilibrium where $\rho_{22}(+\infty)>\rho_{11}(+\infty)$. As a limit, at zero temperature, the quantum fluctuation drives the system evolve to the ground state, where $\rho_{11}(+\infty)=0$ and $\rho_{22}(+\infty)=1$. Such disagreement is caused by the fact that, in Eq. \eqref{eq:incoherent_perturbation}, spontaneous absorptions are not ruled out. In the perturbative quantum field theory \cite{PeskinQFT}, the diagrams in Fig. \ref{fig:feynman}(b) and Fig. \ref{fig:feynman}(d) are omitted, since there are no spontaneous absorptions. To describe the quantum fluctuation, I need to rule out the spontaneous absorptions in the above formulations. An intuitive modification is to set $A^*(\Delta{t},t,\omega,\epsilon)=0$ in the transfer matrix eq. \eqref{eq:transfer_matrix}. However, this modification destroys the unitarity, leading to an unphysical total probability other than $1$. 
			
			In the quantum field theory, reabsorptions like the process in Fig. \ref{fig:feynman}(c), are not ruled out. Simply setting $A^*(\Delta{t},t,\omega,\epsilon)=0$ causes loss of such processes. The reasonable modification is to disable the terms with the spontaneous absorption in Eq. \eqref {eq:monochromatic_shift}, which includes the product $A(\Delta{t}, t, \omega, \epsilon)a_2$ or $A^*(\Delta{t},t,\omega,\epsilon)a^*_2$ (detailed derivations are presented in Appendix \ref{sec:quantum_fluctuation}). Thus the quantum fluctuation with the frequency $\omega$ causes the density matrix shift $\Delta\rho^{\text{QF}}(\Delta{t},t,\omega,\epsilon)$, with the matrix elements
			\begin {subequations}
				\begin {align}
					& \Delta\rho^{\text{QF}}_{11}(\Delta{t},t,\omega,\epsilon) = - \rho_{11}(t)\left|A(\Delta{t},t,\omega,\epsilon)\right|^2, \\
					& \Delta\rho^{\text{QF}}_{12}(\Delta{t},t,\omega,\epsilon) = - \frac{1}{2}\rho_{12}(t)\left|A(\Delta{t},t,\omega,\epsilon)\right|^2, \\
					& \Delta\rho^{\text{QF}}_{21}(\Delta{t},t,\omega,\epsilon) = - \frac{1}{2}\rho_{21}(t)\left|A(\Delta{t},t,\omega,\epsilon)\right|^2, \\
					& \Delta\rho^{\text{QF}}_{22}(\Delta{t},t,\omega,\epsilon) = \rho_{11}(t)\left|A(\Delta{t},t,\omega,\epsilon)\right|^2.
				\end {align}
				\label {eq:nonunitary_transformation}
			\end {subequations}
			By integral over the frequency space, the frequency-independent density matrix shift is 
			\begin {subequations}
				\begin {align}
					& \Delta{\rho}_{11}^{\text{QF}}(\Delta{t},t) = - 2\pi|c|^2n(\Delta{E})\rho_{11}(t)\Delta{t}, \\
					& \Delta{\rho}_{12}^{\text{QF}}(\Delta{t},t) = - \pi|c|^2n(\Delta{E})\rho_{12}(t)\Delta{t}, \\
					& \Delta{\rho}_{21}^{\text{QF}}(\Delta{t},t) = - \pi|c|^2n(\Delta{E})\rho_{21}(t)\Delta{t}, \\
					& \Delta{\rho}_{22}^{\text{QF}}(\Delta{t},t) = 2\pi|c|^2n(\Delta{E}) \rho_{11}(t)\Delta{t}.
				\end {align}
				\label {eq:quantum_fluctuation}
			\end {subequations}
			Eq. \eqref{eq:quantum_fluctuation} shows how the two-level system evolves with the quantum fluctuation. It can be seen that, since
			\begin {equation}
				\Delta\rho^{\text{QF}}_{11}(\Delta{t},t) + \Delta\rho^{\text{QF}}_{22}(\Delta{t},t) = 0,
			\end {equation}
			removing the spontaneous absorption does not lead to an unphysical probability. The deexcitation rate $2\pi|c|^2n(\Delta{E})$ agrees with the Weisskopf-Wigner theory. The decoherence rate $\pi|c|^2n(\Delta{E})$ is half of the deexcitation rate. This result coincide with cases discussed in \cite{Chng2024,Pino2015,Martínez-Martínez_2018}.
			
		\subsection {The thermal fluctuation}
			\label {subsec:thermal_fluctuation}
			
			The thermal fluctuation is not forbidden to excite the ground state to the excited state. Eq. \eqref{eq:energy_mode} indicate that the thermal fluctuation stays in the bosonic equilibrium, the number of perturbation with frequency $\omega$ is expected to be
			\begin {equation}
				{N_\text{B}}(\omega) = \frac{1}{e^{\beta\omega}-1}.
			\end {equation}
			Thus the thermal fluctuation leads to a shift of the density matrix
			\begin {equation}
				\begin {aligned}
				& \Delta\rho^{\text{TF}}(\Delta{t},t) \\
				& \qquad = \lim_{\epsilon\rightarrow0^+} \int_{-\infty}^{+\infty}{d\omega}N_{\text{B}}(\omega)n(\omega)\Delta{\rho}(\Delta{t},t,\omega,\epsilon).
				\end {aligned}
			\end {equation}
 			It can be derived that the elements are
			\begin {subequations}
				\begin {align}
					& \begin {aligned} 
						& \Delta{\rho}^{\text{TF}}_{11}(\Delta{t},t) \\
						& \qquad = - \frac{2\pi|c|^2n(\Delta{E}) (\rho_{11}(t) - \rho_{22}(t)) \Delta{t}}{e^{\beta\omega}-1}, 
					\end {aligned} \\
					& \Delta{\rho}^{\text{TF}}_{12}(\Delta{t},t) = - \frac{2\pi|c|^2n(\Delta{E})\rho_{12}(t)\Delta{t}}{e^{\beta\omega}-1}, \\
					& \Delta{\rho}^{\text{TF}}_{21}(\Delta{t},t) = - \frac{2\pi|c|^2n(\Delta{E})\rho_{21}(t)\Delta{t}}{e^{\beta\omega}-1}, \\
					& \begin {aligned}
						& \Delta{\rho}^{\text{TF}}_{22}(\Delta{t},t) \\
						& \qquad = - \frac{2\pi|c|^2n(\Delta{E})(\rho_{22}(t) - \rho_{11}(t)) \Delta{t}}{e^{\beta\omega}-1}.
					\end {aligned}
				\end {align}
				\label {eq:thermal_fluctuation}
			\end {subequations}
		
		\subsection {A coherent perturbation}
		
			Similar to the case analyzed in Eq. \eqref{eq:monochromatic_shift}, when the two level system is subject to a coherent perturbation, such as a laser beam, which is expressed as
			\begin {equation}
				U(t,\omega) = 
				\begin {pmatrix}
					0 & Ce^{-i\omega{t}} \\
					C^*e^{i\omega{t}} & 0
				\end {pmatrix},
			\end {equation}
			the corresponding shift of the density matrix $\Delta\rho^{\text{CP}}(\Delta{t},t,\omega)$ has elements
			\begin {subequations}
				\begin {align}
					& \begin {aligned}
						\Delta\rho_{11}^{\text{CP}}(\Delta{t},t,\omega) = & -|B(\Delta{t},t,\omega)|^2\rho_{11}(t) \\
						& + iB^*(\Delta{t},t,\omega)\rho_{12}(t) \\
						& - iB(\Delta{t},t,\omega)\rho_{21}(t) \\
						& + |B(\Delta{t},t,\omega)|^2\rho_{22}(t),
					\end {aligned} \\
					& \begin {aligned}
						\Delta\rho_{12}^{\text{CP}}(\Delta{t},t,\omega) = &\ {iB(\Delta{t},t,\omega)}\rho_{11}(t) \\
						& - |B(\Delta{t},t,\omega)|^2\rho_{12}(t) \\
						& + {B^2(\Delta{t},t,\omega)}\rho_{21}(t) \\
						& - {iB(\Delta{t},t,\omega)}\rho_{22}(t),
					\end {aligned} \\
					& 
					\begin {aligned}
						\Delta\rho_{21}^{\text{CP}}(\Delta{t},t,\omega) = & - {iB^*(\Delta{t},t,\omega)}\rho_{11}(t) \\
						& + {{B^*}^2(\Delta{t},t,\omega)}\rho_{12}(t) \\
						& - |B(\Delta{t},t,\omega)|^2\rho_{21}(t) \\
						& + {iB^*(\Delta{t},t,\omega)}\rho_{22}(t),
					\end {aligned} \\
					& 
					\begin {aligned}
						\Delta\rho_{22}^{\text{CP}}(\Delta{t},t,\omega) = &\ |B(\Delta{t},t,\omega)|^2\rho_{11}(t) \\
						& - iB^*(\Delta{t},t,\omega)\rho_{12}(t) \\
						& + iB(\Delta{t},t,\omega)\rho_{21}(t) \\
						& - |B(\Delta{t},t,\omega,)|^2\rho_{22}(t),
					\end {aligned}
				\end {align}
				\label {eq:coherent_shift}
			\end {subequations}
			with
			\begin {equation}
				\begin {aligned}
				B (\Delta{t},t,\omega) &= iCe^{-i(\omega-\Delta{E})t} \Delta{t}.
				\end {aligned}
			\end {equation}
			It is worth noting that keeping $|B(\Delta{t},t,\omega)|^2$, $B^2(\Delta{t},t,\omega)$ and ${B^*}^2(\Delta{t},t,\omega)$ is unnecessary in the limit $\Delta{t}\rightarrow0$. However, these terms help to keep the occupation numbers $\rho_{11}(t)$ and $\rho_{22}(t)$ in the physical interval $[0,1]$ in numerical works with finite $\Delta{t}$. By collecting the results in Eq. \eqref{eq:quantum_fluctuation}, Eq. \eqref{eq:thermal_fluctuation} and Eq. \eqref{eq:coherent_shift}, I arrive at the equation of the evolution of the density matrix
			\begin {equation}
				\begin {aligned}
					\rho(t+\Delta{t}) &= \rho(t) + \Delta\rho^{\text{CP}}(\Delta{t},t,\omega) \\
					&\qquad + \Delta\rho^{\text{QF}}(\Delta{t},t) + \Delta\rho^{\text{TF}}(\Delta{t},t).
				\end {aligned}
				\label {eq:full_evolution}
			\end {equation}
			Dividing both sides of Eq. \eqref{eq:full_evolution} by $\Delta{t}$, I get the optical Bloch equations
			\begin {subequations}
				\begin {align}
					& \begin {aligned}
						\frac{d\rho_{11}(t)}{dt} &= - \frac{2\pi|c|^2n(\Delta{E})e^{\beta\omega}}{e^{\beta\omega}-1}\rho_{11}(t) \\
						&\quad + C^*e^{i(\omega-\Delta{E})t}\rho_{12}(t) \\
						&\quad + Ce^{-i(\omega-\Delta{E})t}\rho_{21}(t) \\
						&\quad + \frac{2\pi|c|^2n(\Delta{E})}{e^{\beta\omega}-1}\rho_{22}(t),
					\end {aligned} \\
					& \begin {aligned}
						\frac{d\rho_{12}(t)}{dt} & = -Ce^{-i(\omega-\Delta{E})t}\rho_{11}(t) \\
						&\quad - \frac{\pi|c|^2n(\Delta{E})(e^{\beta\omega}+1)}{e^{\beta\omega}-1}\rho_{12}(t) \\
						&\quad + Ce^{-i(\omega-\Delta{E})t}\rho_{22}(t),
					\end {aligned} \\
					& \begin {aligned}
						\frac{d\rho_{21}(t)}{dt} & = - C^*e^{i(\omega-\Delta{E})t}\rho_{11}(t) \\
						&\quad - \frac{\pi|c|^2n(\Delta{E})(e^{\beta\omega}+1)}{e^{\beta\omega}-1}\rho_{21}(t) \\
						&\quad + Ce^{i(\omega-\Delta{E})t}\rho_{22}(t),
					\end {aligned} \\
					& \begin {aligned}
						\frac{d\rho_{22}(t)}{dt} & = + \frac{2\pi|c|^2n(\Delta{E})e^{\beta\omega}}{e^{\beta\omega}-1}\rho_{11}(t) \\
						&\quad -C^*e^{i(\omega-\Delta{E})t}\rho_{12}(t) \\
						&\quad -Ce^{-i(\omega-\Delta{E})t}\rho_{21}(t) \\
						&\quad - \frac{2\pi|c|^2n(\Delta{E})}{e^{\beta\omega}-1}\rho_{22}(t).
					\end {aligned}
				\end {align}
			\end {subequations}
		
	\section {Numerical realization}
		\label {sec:results}
		
		Eq. \eqref{eq:full_evolution} can be directly used in the numerical evaluation. With the initial state given by $\rho(0)$, one can iteratively use Eq. \eqref{eq:full_evolution} to obtain the density matrix at any later time $\rho(n\Delta{t})$. In the following, two different conditions are investigated.
		
		\subsection {Spontaneous emission in zero and finite temperatures}
		
\begin {figure} [H]
				\centering
				\includegraphics {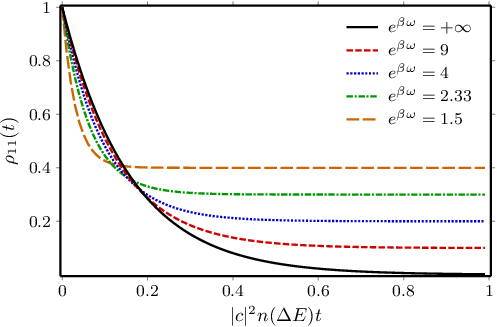}
				\caption {Deexcitation in zero and finite temperature.}
				\label {fig:deexcitation}
			\end {figure}

			The spontaneous emissions are investigated by setting the coherent perturbation shift
			\begin {equation}
				\Delta\rho_{11}^{\text{CP}}(\Delta{t},t,\omega) = 0
			\end {equation} 
			and the initial density matrix
			\begin {equation}
				\rho(0) = \begin {pmatrix}
				1 & 0 \\
				0 & 0
				\end {pmatrix}
			\end {equation}
			in Eq. \eqref{eq:full_evolution}. It is found that with a given temperature, the evolution of the density matrix $\rho(t)$ depends on $|c|^2n(\Delta{E})t$. In Fig. \ref{fig:deexcitation}, numerical evaluations of $\rho_{11}(t)$ are plotted with $e^{\beta\omega}=+\infty, 9, 4, 2.33\ \text{and}\ 1.5$. It is observed that in the corresponding final states, $\rho_{11}(+\infty)$ approaches to the value $0$, $0.1$, $0.2$, $0.3$, $0.4$. This result agrees with the Fermi-Dirac distribution
			\begin {equation}
				N_\text{F}(\Delta{E}) = \frac{1}{e^{\beta\Delta{E}}+1}.
			\end {equation}
			This result indicates that the thermal fluctuation obeying the Bose-Einstein distribution drives the quantum two-level system evolve towards the equilibrium obeying the Fermi-Dirac distribution. One can also observe that in higher temperature, the equilibrations happen quicker.
		
		\subsection {Coherent excitations at finite temperatures}
		
			In this case, the two-level system simultaneously receive the coherence perturbation, the quantum fluctuation and the thermal fluctuation. Suppose at the initial time, the system is in the ground state. The initial condition is
			\begin {equation}
				\rho(0) = \begin {pmatrix}
					0 & 0 \\
					0 & 1
				\end {pmatrix}.
			\end {equation}
			To compare the strength of the coherent perturbation and the fluctuations, I introduce the ratio 
			\begin {equation} 
				\gamma = \frac{|C|^2}{\pi^2|c|^4n^2(\Delta{E})}.
			\end {equation}
			\begin {figure}
				\centering
				\includegraphics {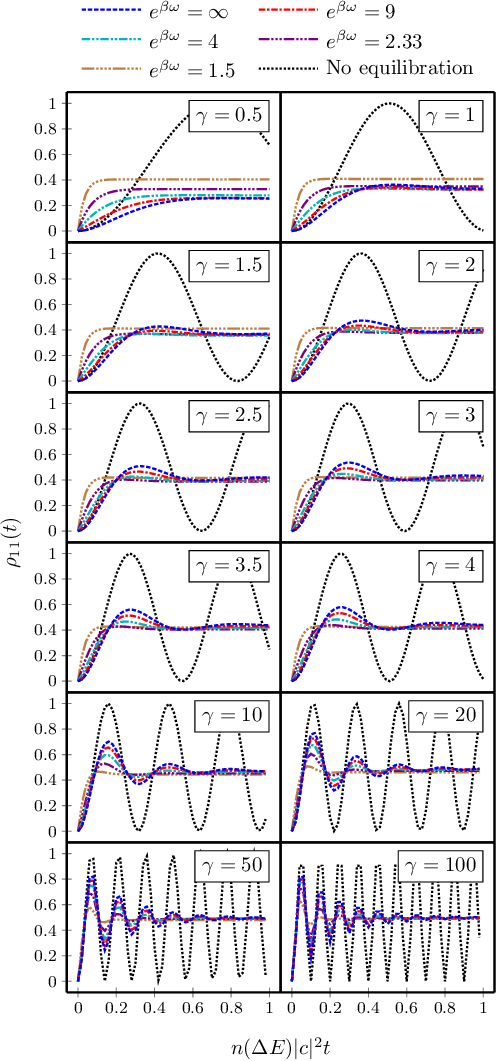}
				\caption {Excitation probability of the coherent perturbation.}
				\label {fig:coherent_excitation}
			\end {figure}
			
			In Fig. \ref{fig:coherent_excitation}, the excitation processes with different $\gamma$ values are plotted. The frequency of the coherent perturbation is set to $\omega=\Delta{E}$. As a comparison, the dotted black line shows the result of pure coherent perturbation, which is obtained by setting $\Delta\rho^{\text{QF}} (\Delta{t}, t) = \Delta\rho^{\text{TF}} (\Delta{t}, t) =0$. Without the fluctuations, the coherent perturbation oscillates the occupation number of the excitation states between $0$ and $1$ with the Rabi frequency. When the quantum and thermal fluctuations are turned on, it can be seen that the oscillation is damped. The damping gets faster in higher temperature. In the final state, the coherent perturbation, the quantum fluctuation and the thermal fluctuation reach a balance. Since $\rho_{11}(t)=1-\rho_{22}(t)$ and $\rho_{12}(t)=\rho^*_{21}(t)$, the stable state can be solved by requiring $\left.\frac{d\rho_{11}(t,\omega)}{dt}\right|_{t=+\infty}=\left.\frac{d\rho_{12}(t,\omega)}{dt}\right|_{t=+\infty}=0$,
			\begin {subequations}
				\begin {align}
					& \begin {aligned}
						& C^*\rho_{12}(+\infty) + C\rho^*_{12}(+\infty) \\
						& \qquad - 2\pi|c|^2n(\Delta{E})\frac{(e^{\beta\omega}+1)\rho_{11}(+\infty)-1}{e^{\beta\omega}-1} = 0,
					\end {aligned} \\
					& \begin {aligned} 
						& -C(2\rho_{11}(+\infty)-1) \\
						& \qquad \qquad - \pi|c|^2n(\Delta{E})\rho_{12}(+\infty)\frac{e^{\beta\omega}+1}{e^{\beta\omega}-1} = 0.
					\end {aligned}
				 \end {align}
				 \label {eq:stable_condition}
			\end {subequations}
			The solution to Eq. \eqref {eq:stable_condition} is
			\begin {equation}
				\rho_{11}(+\infty) = \frac{\gamma\left(e^{\beta\omega}-1\right)^2+2\left(e^{\beta\omega}+1\right)}{2\left[\gamma\left(e^{\beta\omega}-1\right)^2+\left(e^{\beta\omega}+1\right)^2\right]}.
				\label {eq:rho11f}
			\end {equation}
			In real cases, the coherent excitation is much stronger than the spontaneous emissions so that $\gamma$ is usually very large. Eq. \eqref{eq:rho11f} shows that when the coherent perturbation lasts enough long time, the excitation probability approaches $1/2$. Hence the coherent perturbation leads to a stable excitation in the finite temperature field.
					
	\section {Conclusion and outlooks}
		\label {sec:conclusion}
		
		In this article, I derive the optical Bloch equation for a system at finite temperature. The quantum fluctuation and the thermal fluctuation are studied in the framework of quantum mechanics. Density matrix techniques are used to deal with these incoherent perturbations. With the assumption that the quantum fluctuation does not excite the ground state, I derive the shift of the density matrix caused by the quantum fluctuation. Such derivation keeps the unitarity of the probability. The deexcitation rate predicted by this method agrees with the Weisskopf-Wigner theory. It also shows the decay of the coherence which agrees with the analysis in various environments. To include the finite-temperature effects, the thermal fluctuation is introduced as the incoherence perturbation with the Bose-Einstein distribution. It is shown that the quantum fluctuation and the thermal fluctuation drives the quantum system to the equilibrium, which follows the Fermi-Dirac distribution. By combining the above result with the result of the coherent perturbation, it is shown that the quantum fluctuation and the thermal fluctuation change the long-time behavior of the excitation by the coherence perturbation.
		
		An assumption is used throughout the discussion in this article - the fluctuations are not changed during the back and forth emission and absorption processes. This is true for the electric magnetic fluctuations in the free space. It also keeps to be a good approximation over a wide range of operating conditions. To immediately demonstrate the effects of the quantum fluctuation and the thermal fluctuation, in this article, a non-degenerate two-level system is discussed. This is a simplification of real cases. However, with the increase of number of states, the derivation of the optical Bloch equation becomes complicated. This problem may be solved by improving the algorithm. The discussions in this article is restricted to the one-body system. It is useful but challenging to generalize the equations for few- and many-body fermionic systems, where the Pauli exclusion puts restrictions on the transitions.
		
	\appendix
	
	\section {Zero-dimensional Weisskopf-Wigner theory}
		\label {sec:W-W-theory}
	
		Here we follow the Weisskopf-Wigner theory to derivate the spontaneous deexcitation rate. Unlike the original derivation in \cite {Weisskopf1930}, to compare with the method discussed in the article, a zero-dimensional model is considered, with the Hamiltonian
		\begin {equation}
			\begin {aligned}
				H_{\text{W.W.}} =\ & \omega |1\rangle\langle1|\\
				& \begin {aligned}
				+ &\ \int d\tilde{\omega} \left[ \tilde{\omega} a^\dagger(\tilde{\omega}) a(\tilde{\omega}) \right. \\
				& \left. + g(\tilde{\omega})|1\rangle\langle2|a(\tilde{\omega}) \right. \\
				& \left. + g^*(\tilde{\omega})|2\rangle\langle1|a^{\dagger}(\omega)\right],
				\end {aligned}
			\end {aligned}
		\end {equation}
		Noting that $\langle\gamma(\omega_1)|\gamma(\omega_2)\rangle=\delta(\omega_1-\omega_2)$, the operators $a^\dagger(\tilde{\omega})$ and $a(\tilde{\omega})$ have the dimension $[E]^{-1/2}$, the coupling constants $g(\tilde{\omega})$ and $g^*(\tilde{\omega})$ have the dimension $[E]^{1/2}$. For this Hamiltonian, the basis are divided into two sets, one contains the single particle in the excited state $|1\rangle$, which has the energy $\omega$. The other is the ground state particle with a photon $|2,\gamma(\tilde{\omega}) \rangle$. Thus the ansatz is
		\begin {equation}
			\begin {aligned}
				|\Psi(t)\rangle = \ & b_1(t)e^{-i\omega{t}}|1\rangle \\
				& \hspace{-3em} + \int d\tilde{\omega} b_2(t,\tilde{\omega})e^{-i\tilde{\omega}t}|2,\gamma(\tilde{\omega})\rangle,
			\end {aligned}
		\end {equation}
		where $b_1(t)$ is dimensionless and $b_2(t,\tilde{\omega})$ has the dimension $[E]^{1/2}$. With the TDSE leads to
		\begin {subequations}
			\begin {align}
				& 
				\begin {aligned}
					i\frac{\partial}{\partial{t}}b_1(t) =& \int d\tilde{\omega}g(\tilde{\omega}) \\
					& \hspace{2ex} \times b_2(t,\tilde{\omega})e^{-i(\tilde{\omega}-\omega)t} \\
				\end {aligned} 
				\label {eq:C1}
				\\
				& i\frac{\partial}{\partial{t}}b_2(t,\tilde{\omega}) = g^*(\tilde{\omega})b_1(t)e^{-i(\omega-\tilde{\omega})t}
				\label {eq:C2}
			\end {align}
		\end {subequations}
		Eq. \eqref{eq:C2} leads to
		\begin {equation}
			b_2(t,\tilde{\omega}) = -ib_1(t)g^*(\tilde{\omega})\int_0^tdte^{-i(\omega-\tilde{\omega})t},
			\label {eq:C2solution}
		\end {equation}
		where a Markovian approximation is used so that $b_1(t)$ is taken out of the integral. In the limit $t\rightarrow+\infty$
		\begin {equation}
			\begin {aligned}
				\lim_{t\rightarrow+\infty} b_2(t,\tilde{\omega}) =\ & -i b_1(t)g^*(\tilde{\omega}) \\
				& \hspace{-10ex} \times \left[\pi\delta(\omega-\tilde{\omega})-iP\left(\frac{1}{\omega-\tilde{\omega}}\right)\right]
			\end {aligned}
		\end {equation}
		By omitting the Cauchy principal part $P\left(\frac{1}{\omega-\tilde{\omega}}\right)$, one obtains
		\begin {equation}
			\begin {aligned}
				i\frac{\partial}{\partial{t}}b_1(t) =\ & -i\pi \int d\tilde{\omega} |g(\tilde{\omega})|^2 \\
				&\times b_1(t) \delta(\omega-\tilde{\omega}) e^{-i(\tilde{\omega}-\omega)t} \\
				=\ & - i\pi |g(\omega)|^2 b_1(t). \\
			\end {aligned}
		\end {equation}
		This leads to the solution
		\begin {equation}
			b_1(t) = e^{-\pi|g(\omega)|^2t},
		\end {equation}
		where leads to the probability
		\begin {equation}
			P_1(t) = e^{-2\pi|g(\omega)|^2t}.
		\end {equation}
	
	\section {Integrations with the residue theorem}
		\label {sec:integration}
	
\begin {figure} [H]
			\includegraphics [width=8.5cm] {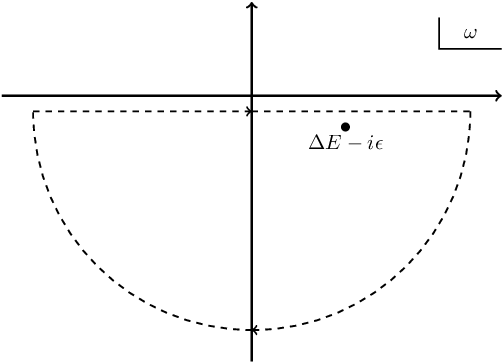}
			\caption {The integral path for $I_1$.}
			\label {fig:contour}
		\end {figure}

		The integral
		\begin {equation}
			I[\omega,\Delta{E},t,f(\omega)] = \int_{-\infty}^{+\infty}d\omega f(\omega)\frac{e^{-i(\omega-\Delta{E})t}}{\omega-\Delta{E}+i\epsilon}
			\label {eq:integral_I}
		\end {equation}
		has a pole at $\omega=\Delta{E}-i\epsilon$. Assume $f(\omega)e^{-i(\omega-\Delta{E})t}$ vanishes when $\text{Im}(\omega)=-\infty$ (since $f(\omega)=n(\omega)$ in the cases of quantum fluctuations and $f(n)=n(\omega)/(e^{\beta\omega}-1)$ in the cases of thermal fluctuations, this assumption is always valid within the discussion in the article), the integration Eq. \eqref{eq:integral_I} is converted into
		\begin {equation}
			I[\omega,\Delta{E},t,f(\omega)] = \oint_{\mathcal{C}}d\omega f(\omega)\frac{e^{-i(\omega-\Delta{E})t}}{\omega-\Delta{E}+i\epsilon},
		\end {equation}
		where the contour $\mathcal{C}$ is plotted in Fig. \ref{fig:contour}. According to the residue theorem,
		\begin {equation}
			\begin {aligned}
				&\quad I [\omega,\Delta{E}, t, f(\omega)] \\
				&= - \left.2\pi{i} f(\omega) e^{-i(\omega-\Delta{E})t}\right|_{\omega=\Delta{E}-i\epsilon} \\
				&= - 2\pi i f(\Delta{E}-i\epsilon) e^{-\epsilon{t}}.
			\end {aligned}
		\end {equation}
		With this result, I can obtain that in the limit $\epsilon\rightarrow0^+$,
		\begin {equation}
			\begin {aligned}
				&\phantom{=}\lim_{\epsilon\rightarrow0^+} \int_{-\infty}^{+\infty}d\omega{n}(\omega)A(\Delta{t},t,\omega,\epsilon) \\
				&= - \lim_{\epsilon\rightarrow0^+} c\int_{-\infty}^{+\infty}d\omega{n}(\omega)\frac{e^{-i(\omega-\Delta{E})(t+\Delta{t})}}{\omega-\Delta{E}+i\epsilon} \\
				&\qquad + \lim_{\epsilon\rightarrow0^+} c\int_{-\infty}^{+\infty}d\omega{n}(\omega)\frac{e^{-i(\omega-\Delta{E})t}}{\omega-\Delta{E}+i\epsilon} \\
				&= - c\lim_{\epsilon\rightarrow0^+}I[\omega,\Delta{E},t+\Delta{t},n(\omega)] \\
				&\qquad + c\lim_{\epsilon\rightarrow0^+}I[\omega,\Delta{E},t,n(\omega)] \\
				&= 0.
			\end {aligned}
			\label {eq:int_A}
		\end {equation}
		The complex conjugate of Eq. \eqref{eq:int_A}
		\begin {equation}
			\begin {aligned}
				&\phantom{=}\lim_{\epsilon\rightarrow0^+} \int_{-\infty}^{+\infty}d\omega{n}(\omega)A^*(\Delta{t},t,\omega,\epsilon) = 0.
			\end {aligned}
			\label {eq:int_A*}
		\end {equation}
		Since, there are no terms of power $-1$, the integration of the second-order term		
		\begin {equation}
			\begin {aligned}
				&\phantom{=}\lim_{\epsilon\rightarrow0^+} \int_{-\infty}^{+\infty}d\omega{n}(\omega)A^2(\Delta{t},t,\omega,\epsilon) \\
				&= \lim_{\epsilon\rightarrow0^+} c^2\int_{-\infty}^{+\infty}d\omega{n}(\omega) \\
				&\qquad \frac{\left[e^{-i(E_2+\omega-E_1)(t+\Delta{t})}-e^{-i(E_2+\omega-E_1)t}\right]^2}{(E_2+\omega-E_1+i\epsilon)^2} \\
				&= 0
			\end {aligned}
			\label {eq:int_A2}
		\end {equation}
		and
		\begin {equation}
			\begin {aligned}
				\phantom{=}\lim_{\epsilon\rightarrow0^+} \int_{-\infty}^{+\infty}d\omega{n}(\omega){A^*}^2(\Delta{t},t,\omega,\epsilon) = 0.
			\end {aligned}
			\label {eq:int_A*2}
		\end {equation}
		The only contributing term is
		\begin {equation}
			\begin {aligned}
				&\phantom{=} \lim_{\epsilon\rightarrow0^+} \int_{-\infty}^{+\infty}d\omega{n}(\omega)|A(\Delta{t},t,\omega,\epsilon)|^2 \\
				&= \lim_{\epsilon\rightarrow0^+}|c|^2\int_{-\infty}^{+\infty}d\omega{n}(\omega) \\
				&\qquad \times \frac{e^{-i(\omega-\Delta{E})(t+\Delta{t})}-e^{-i(\omega-\Delta{E})t}}{\omega-\Delta{E}+i\epsilon} \\
				&\qquad\qquad \times\frac{e^{i(\omega-\Delta{E})(t+\Delta{t})}-e^{i(\omega-\Delta{E})t}}{\omega-\Delta{E}-i\epsilon} \\
				&= \lim_{\epsilon\rightarrow0^+} |c|^2 \int_{-\infty}^{+\infty} \frac{d\omega n(\omega)}{(\omega-\Delta{E}+i\epsilon)(\omega-\Delta{E}-i\epsilon)} \\
				&\qquad \times \left[ 2 - e^{i(\omega-\Delta{E})\Delta{t}} - e^{-i(\omega-\Delta{E})\Delta{t}} \right] \\
				&= \lim_{\epsilon\rightarrow0^+} |c|^2 I[\omega, \Delta{E}, 0, \frac{2n(\omega)}{\omega-\Delta{E}-i\epsilon}] \\
				&\qquad - \lim_{\epsilon\rightarrow0^+} |c|^2 I[\omega, \Delta{E}, 0, \frac{n(\omega )e^{i(\omega-\Delta{E})\Delta{t}}}{\omega-\Delta{E}-i\epsilon}] \\
				&\qquad\quad - \lim_{\epsilon\rightarrow0^+} |c|^2 I[\omega, \Delta{E}, 0, \frac{n(\omega)e^{-i(\omega-\Delta{E})\Delta{t}}}{\omega-\Delta{E}-i\epsilon}] \\
				&= 2\pi|c|^2n(\Delta{E}) \Delta{t}.
			\end {aligned}
		\end {equation}
		
	\section {Term by term analysis of the quantum fluctuation}
		\label {sec:quantum_fluctuation}
		
		In Appendix \ref{sec:integration}, it is shown that only the terms with $|A(\Delta{t},t,\omega,\epsilon)|$ contribute to the time derivative of $\rho(t)$. Thus, to derive the shift by the quantum fluctuation, it is necessary to expand Eq. \eqref {eq:density_evolution} to the second order. Then the terms include spontaneous absorptions are manually removed in the derivation.
		\begin {equation}
					\begin {aligned} 
						& \Delta\rho^{\text{QF}}_{11}(\Delta{t},t,\omega,\epsilon) \\
						=\ & \frac{1}{4} \left[\left({A^*(\Delta{t},t,\omega,\epsilon)A(\Delta{t},t,\omega,\epsilon)}-2\right)a^*_1(t)\vphantom{\xcancel{A}} \right. \\
						& \qquad \left. -\xcancel{2iA^*(\Delta{t},t,\omega,\epsilon)a^*_2(t)}\right] \\
						& \times \left[\xcancel{2iA(\Delta{t},t,\omega,\epsilon)a_2(t)} \right. \\
						& \qquad \left. + \vphantom{\tilde{A}} a_1(t)\left(A(\Delta{t},t,\omega,\epsilon)A^*(\Delta{t},t,\omega,\epsilon)-2\right)\right] \\
						& - \rho_{11}(t) \\
						=\ & \frac{1}{4} \rho_{11}(t) \left( - 4 |A(\Delta{t},t,\omega,\epsilon)|^2 + 4 \right) - \rho_{11}(t) \\
						=\ & - \rho_{11}(t) \left|A(\Delta{t},t,\omega,\epsilon)\right|^2,
					\end {aligned} \\
		\end {equation}
		\begin {equation}
					\begin {aligned} 
						& \Delta\rho^{\text{QF}}_{12}(\Delta{t},t,\omega,\epsilon) \\
						=\ & \frac{1}{4} \left[\left(\xcancel{A^*(\Delta{t},t,\omega,\epsilon)A(\Delta{t},t,\omega,\epsilon)}-2\right)a_2^*(t) \right. \\
						& \qquad \left.-2iA(\Delta{t},t,\omega,\epsilon)a_1^*(t)\vphantom{\tilde{A^*}}\right] \\
						& \times \left[\xcancel{2iA(\Delta{t},t,\omega,\epsilon)a_2(t)} \right. \\
						& \qquad \left. + \vphantom{\tilde{A}}a_1(t)\left(A(\Delta{t},t,\omega,\epsilon)A^*(\Delta{t},t,\omega,\epsilon)-2\right)\right] \\
						& - \rho_{12}(t) \\
						=\ & \frac{1}{4} \left[-2\rho_{12}(t)(|A(\Delta{t},t,\omega,\epsilon)|^2-2)\right] - \rho_{12}(t) \\
						=\ & - \frac{1}{2}\rho_{12}|A(\Delta{t},t,\omega,\epsilon)|^2,
					\end {aligned} \\
		\end {equation}
		\begin {equation}
			\begin {aligned} 
				& \Delta\rho_{21}^{\text{QF}}(\Delta{t},t,\omega,\epsilon) \\
				=\ & \frac{1}{4} \left[\left(A^*(\Delta{t},t,\omega,\epsilon)A(\Delta{t},t,\omega,\epsilon)-2\right)a_1^*(t) \vphantom{-\xcancel{2iA^*(\Delta{t},t,\omega,\epsilon)a_2^*(t)}} \right. \\
				& \qquad \left. -\xcancel{2iA^*(\Delta{t},t,\omega,\epsilon)a_2^*(t)}\right]  \\
				& \times \left[2iA^*(\Delta{t},t,\omega,\epsilon)a_1(t) \vphantom{+a_2(t)\left(\xcancel{A(\Delta{t},t,\omega,\epsilon)A^*(\Delta{t},t,\omega,\epsilon)}-2\right)} \right. \\
				& \quad \left. +a_2(t)\left(\xcancel{A(\Delta{t},t,\omega,\epsilon)A^*(\Delta{t},t,\omega,\epsilon)}-2\right)\right] \\
				& - \rho_{21}(t) \\
				=\ & \frac{1}{4} [-2\rho_{21}(|A(\Delta{t},t,\omega,\epsilon)|^2-2)] - \rho_{21}(t) \\
				=\ & -\frac{1}{2}\rho_{21}|A(\Delta{t},t,\omega,\epsilon)|^2,
			\end {aligned} \\
		\end {equation}
		\begin {equation}
			\begin {aligned} 
				& \Delta\rho^{\text{QF}}_{22}(\Delta{t},t,\omega,\epsilon) \\
				=\ & \frac{1}{4} \left[\left(\xcancel{A^*(\Delta{t},t,\omega,\epsilon)A(\Delta{t},t,\omega,\epsilon)} -2\right)a_2^*(t) \right. \\
				& \qquad \left. -2iA(\Delta{t},t,\omega,\epsilon)a_1^*(t) \vphantom{\left(\xcancel{A^*(\Delta{t},t,\omega,\epsilon)A(\Delta{t},t,\omega,\epsilon)} -2\right)a_2^*(t)} \right] \\
				& \times \left[2iA^*(\Delta{t},t,\omega,\epsilon)a_1(t) \vphantom{+a_2(t)\left(\xcancel{A(\Delta{t},t,\omega,\epsilon)A^*(\Delta{t},t,\omega,\epsilon)}-2\right)} \right. \\
				& \quad \left. +a_2(t)\left(\xcancel{A(\Delta{t},t,\omega,\epsilon)A^*(\Delta{t},t,\omega,\epsilon)}-2\right)\right] \\
				=\ & \frac{1}{4} \left(4 \rho_{11}(t)|A(\Delta{t},t,\omega,\epsilon)|^2 \right. \\
				\ & \left.-4i\rho_{12}A^*(\Delta{t},t,\omega,\epsilon)+ 4i\rho_{21}A(\Delta{t},t,\omega,\epsilon) \right. \\
				\ & \left. +4\rho_{22}(t) \vphantom{A^2} \right) - \rho_{22}(t) \\
				=\ & \rho_{11}(t) |A(\Delta{t}, t, \omega, \epsilon)|^2 - i\rho_{12}A^*(\Delta{t},t,\omega,\epsilon) \\
				& +i\rho_{21}(t) A^*(\Delta{t},t,\omega,\epsilon).
			\end {aligned}
		\end {equation}
			In Eq. \eqref {eq:nonunitary_transformation}, the cancellation $A(\Delta{t},t,\epsilon,\omega)a_2(t)$ and $A^*(\Delta{t},t,\epsilon,\omega)a^*_2(t)$ are removed manually. Using the result in Appendix \ref{sec:integration}, one can obtain Eq. \eqref {eq:quantum_fluctuation}.

	\bibliography {all}
	\bibliographystyle {unsrt}

\end {document}